\documentclass[preprint]{aastex}
\shorttitle{SZ Herculis: Physical Nature and Orbital Behavior}
\shortauthors{Lee et al.}

\begin{document}

\title {The Algol System SZ Herculis: Physical Nature and Orbital Behavior}
\author{Jae Woo Lee, Chung-Uk Lee, Seung-Lee Kim, Ho-Il Kim, and Jang-Ho Park}
\affil{Korea Astronomy and Space Science Institute, Daejon 305-348, Korea}
\email{jwlee@kasi.re.kr, leecu@kasi.re.kr, slkim@kasi.re.kr, hikim@kasi.re.kr, pooh107162@kasi.re.kr}

\begin{abstract}
Multiband CCD photometric observations of SZ Her were obtained between 2008 February and May. The light curve was completely
covered and indicated a significant temperature difference between both components. The light-curve synthesis presented in
this paper indicates that the eclipsing binary is a classical Algol-type system with parameters of $q$=0.472, $i$=87$^\circ$.57, 
and $\Delta$($T_{1}$--$T_{2}$)=2,381 K; the primary component fills approximately 77\% of its limiting lobe and is slightly 
larger than the lobe-filling secondary. More than 1,100 times of minimum light spanning more than one century were used to 
study an orbital behavior of the binary system. It was found that the orbital period of SZ Her has varied due to a combination 
of two periodic variations with cycle lengths of $P_3$=85.8 yr and $P_4$=42.5 yr and semi-amplitudes of $K_3$=0.013 d and 
$K_4$=0.007 d, respectively. The most reasonable explanation for them is a pair of light-time effects (LITEs) driven 
by the possible existence of two M-type companions with minimum masses of $M_3$=0.22 M$_\odot$ and $M_4$=0.19 M$_\odot$,
that are located close to the 2:1 mean motion resonance. If two additional bodies exist, then the overall dynamics of 
the multiple system may provide a significant clue to the formation and evolution of the eclipsing pair.
\end{abstract}

{}

\section{INTRODUCTION}

SZ Her ($\rm BD+33^{o} 2930$, GSC 2610-1209, HIP 86430, TYC 2610-1209-1) is an Algol-type system with an orbital period of 0.818 d 
and was announced to be a variable by Ceraski (1908) and also Dun\'er et al. (1909). Although the first observations of 
the system date back to 1902 (Shapley 1913; Russell \& Shapley 1914; Dugan 1923), its properties are poorly known compared to 
those of other short-period Algols. To date in the published literature, only one light-curve analysis has been published and 
it was presented by Giuricin \& Mardirossian (1981). They analyzed the two-color photoelectric light curves of Broglia et al. (1955) 
using the WINK model (Wood 1972) and concluded that the system is a semi-detached Algol-type binary with a mass ratio of $q$=0.4, 
an orbital inclination of $i$=87$^\circ$.9, and a temperature difference between the components of $\Delta T$=2,600 K. Recently, 
Sz\'ekely (2003) and Dvorak (2009) performed CCD observations in order to locate $\delta$ Scuti-type pulsations but failed 
to detect them.

Although the orbital period of SZ Her has been examined several times (Kreiner 1971; Mallama 1980; Zavala et al. 2002), 
a detailed study of its orbital period was made by Sz\'ekely (2003) and Soydugan (2008). They reported that the period change 
can be described using either a sine curve or a single light-time effect (LITE) ephemeris due to a third body with 
implied periods of 66 and 71 yr, respectively. Soydugan (2008) also suggested that the timing residuals from the LITE fit 
indicate an additional short-term oscillation with a period within about 20 yr. More than one thousand eclipse timings, 
spanning $\sim$ 110 yr, should be sufficient to resolve the confusion regarding the orbital behavior of SZ Her. Nonetheless, 
the period variation of this system has not yet been studied as conclusively as required.  In this article, 
a new photometric study of SZ Her based on modern observations and analyses is presented, and it is demonstrated that 
the SZ Her system is likely a quadruple one containing two low-mass companions.

\section{NEW OBSERVATIONS}

We performed new CCD photometry of SZ Her on 13 nights from 2008 February 28 through May 17. The observations were taken with 
a SITe 2K CCD camera and a $BVRI$ filter set attached to the 61-cm reflector at Sobaeksan Optical Astronomy Observatory (SOAO) 
in Korea. The instrument and reduction method used were the same as those described by Lee et al. (2007, 2010b). GSC 2610-1116 
($\rm BD+33^{o} 2925$, TYC 2610-1116-1) and GSC 2610-0821, imaged on the chip at the same time as the program target, were 
selected as comparison and check stars, respectively. The 1$\sigma$-values of the dispersions of the magnitude differences 
between these stars are within $\pm$0.01 mag for all bandpasses. The coordinates and Tycho magnitudes for the three stars of interest 
are given in Table 1. A total of 1,728 individual observations were obtained among the four bandpasses (435 in $B$, 437 in $V$, 
439 in $R$, and 417 in $I$) and a sample of them is listed in Table 2. The light curves of SZ Her defined by the SOAO observations 
are plotted in Figure 1 as the (V$-$C) differential magnitudes {\it versus} orbital phase, which was computed according to 
the ephemeris for our binary model determined later in this article with the Wilson-Devinney synthesis code 
(Wilson \& Devinney 1971, hereafter W-D). 

In addition to these complete light curves, two eclipse timings were observed in 2004 June and 2011 May using the same telescope. 
The 2004 data were collected using the SITe 2K CCD camera and $B$ filter, and the 2011 ones using an FLI IMG4301E CCD camera and 
$BV$ filters. GSC 2610-1116 also served as the comparison star for these data collections. Details of the new detector have been 
given previously by Lee et al. (2011).

\section{LIGHT-CURVE SYNTHESIS AND ABSOLUTE DIMENSIONS}

The shape of the light curve of SZ Her is very similar to that of Algol type. As shown in Figure 1, the light curve was completely
covered and the depth differences between the primary and secondary eclipses indicate a significant temperature difference 
between the two components. In order to understand the physical properties of the system, the $BVRI$ light curves in this study 
were analyzed simultaneously in a manner almost identical to those for XX Cep (Lee et al. 2007) and CL Aur (Lee et al. 2010a) 
using the 2003 version\footnote {ftp://ftp.astro.ufl.edu/pub/wilson/} of the W-D code and the so-called $q$-search procedure 
(cf. Lee et al. 2008).  In this paper, the subscripts 1 and 2 refer to the primary and secondary stars being eclipsed at Min I 
(at phase 0.0) and Min II, respectively. 

The surface temperature of the hotter, and presumably more massive, primary star was assumed to be $T_{1}$=7,270 K from Flower's 
(1996) table, according to ($B-V$)=$+$0.323$\pm$0.041 in the Tycho-2 Catalog (H\o g et al. 2000) and $E$($B-V$)=$+$0.035 
calculated following Schlegel et al. (1998). The gravity-darkening exponents were initialized at standard values of $g_1$=1.0 
and $g_2$=0.32 and the bolometric albedos at $A_1$=1.0 and $A_2$=0.5, as surmised from the components' temperatures. 
Linear bolometric and monochromatic limb-darkening coefficients were interpolated from the values of van Hamme (1993) in 
concert with the model atmosphere option. Furthermore, a synchronous rotation for both components and a circular orbit were adopted 
and the detailed reflection effect was considered.

The only photometric solution of SZ Her was reported by Giuricin \& Mardirossian (1981) 30 years ago and a spectroscopic orbit has 
not yet been established.  Thus, an extensive $q$-search procedure was conducted for a series of models with varying $q$ values.
In this process, we first considered the orbital inclination ($i$), effective temperature ($T$), 
dimensionless surface potential ($\Omega$), and luminosity ($L_1$). This procedure showed acceptable photometric solutions for 
mode 5 only, which are semi-detached systems for which the less massive secondary stars accurately fill their inner Roche lobes.
As displayed in Figure 2, the $q$-search results indicate that the minimum value of the weighted sum of the squared residuals 
($\Sigma$) is approximately $q$=0.49.  Then, we treated this $q$ value as an adjustable parameters and included limb-darkening 
coefficients, albedos, and gravity darkening exponents as additional free variables. The final values are given in Table 3 and 
are plotted in Figure 1 as solid curves. In the figure, the model light curves describe the SOAO multiband data satisfactorily. 
Our light-curve synthesis demonstrates that SZ Her is an Algol-type semi-detached system in which the primary component fills 
its limiting lobe by approximately 77\% and is slightly larger than the lobe-filling secondary component. 
The gravity darkening exponent of the secondary component is consistent with the standard convective $g$ value, while its albedo 
is close to the standard radiative $A$ value.  In these analyses, we searched for a possible third light source but found that 
the parameter remained zero within its error. 

The dereddened color ($B-V$)$_0$=$+$0.29 and temperature of the primary component correspond to a normal main-sequence star with 
a spectral type of about A9. We estimated the absolute dimensions for the binary system from our photometric solution and 
from Harmanec's (1988) relation between the spectral type and stellar mass.  These are given in Table 4, where the luminosity ($L$) 
and bolometric magnitudes ($M_{\rm bol}$) were computed by adopting $T_{\rm eff}$$_\odot$=5,780 K and $M_{\rm bol}$$_\odot$=+4.73 
for solar values. For the absolute visual magnitudes ($M_{\rm V}$), we used the bolometric corrections (BCs) appropriate for 
the temperature of each component from the expression between $\log T$ and BC given by Torres (2010). With an apparent visual magnitude
of $V$=+10.06 and the interstellar absorption of $A_{\rm V}$=0.11, we calculated the distance of the system to be 294 pc. This result 
is consistent with 306 pc taken from the trigonometric parallax (3.27$\pm$1.09 mas; Perryman et al. 1997).  

A comparison of the SZ Her parameters with the mass-radius, mass-luminosity, and Hertzsprung-Russell (HR) diagrams 
(\. Ibano\v{g}lu et al. 2006) clearly demonstrates that the primary component lies in the main-sequence band, while 
the secondary is slightly beyond the terminal-age main sequence and its radius and luminosity are about two times oversized 
and more than four times overluminous compared with dwarf stars of the same mass. In these diagrams, the locations of 
the two components conform to the general pattern of classical Algols. The mass and temperature of the secondary star 
correspond to a spectral type of approximately K2 to K3.

\section{ORBITAL PERIOD STUDY}

From the current observations, six new times of minimum light and their errors were determined using the method of 
Kwee \& van Woerden (1956) and with the weighted mean for the values in each filter. These are listed in Table 5, wherein 
73 additional eclipses were obtained using the data from the WASP (Wide Angle Search for Planets) public archive (Butters et al. 2010). 
For ephemeris computations, we have collected a total of 1050 timings (949 visual, 20 photographic, 16 photoelectric and 65 CCD) 
from the literature (Kreiner et al. 2001; Baldwin \& Samolyk 2002, 2004; Locher 2002a, 2002b; Baki\c s et al. 2003; Sz\'ekely 2003; 
Diethelm 2003, 2004; Nelson 2005; Cook et al. 2005; Kim et al. 2006; Nagai 2004, 2006; H\"ubscher, et al. 2006, 2009; 
Senavci et al. 2007; Samolyk 2008a, 2008b; Br\'at et al. 2008; Liakos \& Niarchos 2009; Do\u{g}ru et al. 2009, 2011; Dvorak 2010; 
Erkan et al. 2010; H\"ubscher \& Monninger 2011) to add to the current measurements. Most earlier timings were extracted from 
the database published by Kreiner et al. (2001).  The secondary minima are much shallower than the primary ones and 
the $O$--$C$ residuals from the two eclipse types are in phase with each other. Thus, we did not use all secondary eclipses in 
the subsequent analysis. Because many timings have been published without errors, the following standard deviations were assigned 
to the timing residuals based on an observational technique: $\pm$0.0036 d for visual, $\pm$0.0020 d for photographic, and 
$\pm$0.0013 d for photoelectric and CCD minima. Relative weights were then scaled from the inverse squares of these values 
(Lee et al. 2007). 

Previous researchers (Sz\'ekely 2003; Soydugan 2008) have suggested that the period variations of SZ Her can be represented using 
an LITE caused by the presence of a third body in the system. First of all, we fitted the minimum epochs to 
the single LITE ephemeris as follows:
\begin{eqnarray}
C_1 = T_0 + PE + \tau_3,
\end{eqnarray}
where $\tau_{3}$ is the LITE due to a hypothetical distant companion to the eclipsing close pair (Irwin 1952, 1959) and includes 
five parameters ($a_{12}\sin i_3$, $e$, $\omega$, $n$, $T$). The Levenberg$-$Marquardt (LM) technique (Press et al. 1992) was 
used to evaluate the seven parameters of the ephemeris. The results are summarized in column (2) of Table 6, together with 
their related quantities. As displayed in Figure 3, the single LITE ephemeris failed to provide a satisfactory result. 

Because the timing residuals in the lower panel of the figure indicate the existence of further effects, some combination of long- 
and short-term period variations appears possible. Using the PERIOD04 program (Lenz \& Breger 2005), 
which can extract individual frequencies from the multi-periodic content of an astronomical time series containing gaps, we looked 
to see if the $O$--$C_{1}$ residuals from the linear terms represent multi-periodic variations. As can be seen from Figure 4, 
two frequencies of $f_1$=0.0000353 cycle d$^{-1}$ and $f_2$=0.0000706 cycle d$^{-1}$ were detected corresponding to 77.6 yr and 
38.8 yr, respectively. Therefore, after considering the two periods and testing several other forms, such as 
a quadratic {\it plus} single-LITE ephemeris, a two-LITE ephemeris and a quadratic {\it plus} two-LITE ephemeris, we found that 
the times of minimum light are best fitted using the following ephemeris: 
\begin{eqnarray}
C_2 = T_0 + PE + \tau_3 + \tau_4.
\end{eqnarray}
where $\tau_{3}$ and $\tau_{4}$ are the light times due to a third and fourth body, respectively. The LM method was 
applied again in order to simultaneously locate the LITE parameters of the third and newly assumed fourth bodies. The calculations 
converged quickly to yield the entries listed in columns (3)--(4) of Table 6.  The $O$--$C_{2}$ residuals from the linear terms 
are plotted in the top panel of Figure 5. The second and third panels display the $\tau_3$ and $\tau_4$ orbits, respectively, 
and the bottom panel represents the residuals from the full ephemeris. The long-term orbit ($\tau_{3}$) are currently 
preliminary because about 1.3 cycles of the 86-yr period have been covered, while the short-term orbit ($\tau_{4}$) has 
a relatively high determinacy because the observations have already covered about 2.6 cycles. 

On the other hand, it is alternatively possible that the $O$--$C$ diagrams may be described by abrupt period changes instead of
continuous period variations. As displayed in Figures 3 and 5, the orbital period of SZ Her seemed to experience period jumps 
around years 1920, 1960, 1978, 1987, 2002, and 2008. They could possibly have been produced either by episodic mass transfer events 
or by impulsive mass ejections from one (or both) component(s). Assuming constant periods before and after the years, 
we applied linear least-squares fits separately to the seven sections. The results are plotted as the thick lines in the top panel 
of Figure 5. A combination of the straight lines resulted in a larger $\chi^2$ = 1.031 than the two-LITE ephemeris. As seen in 
the figure, the sudden period changes seem to have alternated cyclically in algebraic sign. Then, the alternating sign changes 
would require some preferred reciprocating mechanism. However, in view of the semi-detached nature of the binary, no pair of 
unstable locales is obvious and then it is difficult for the jumps to produce perfectly smooth and tilted periodic components 
in the $O$--$C$ residuals. These might be an indication of sinusoidal variations rather than abrupt period changes.

\section{DISCUSSION}

For the eclipsing binary SZ Her we obtained six times of minimum light from the eclipse light curves using the 61cm-reflector at SOAO.
This data set was further augmented with additional photometric data provided by the WASP public archive. These two data sets 
were added to previous measurements of minimum light from earlier epochs. We then carried out a detailed analysis of 
the resulting $O$--$C$ diagram by fitting Keplerian LITE models to the data set. The best fit to the data suggests the existence 
of two companions with the orbital parameters listed in Table 6. The period ratio of $P_{3}$/$P_{4}$=2.02$\pm$0.06 would 
suggest that the two companions are in a 2:1 mean motion orbital resonance. We think that a long-term gravitational interaction 
between the two objects would result in capture into the 2:1 resonant configuration (cf. Kley et al. 2004). To our knowledge, 
this would be the fourth case when two circumbinary companions would be in or close to any kind of resonance.  Lee et al. (2009) 
discovered two substellar companions revolving around the sdB+M eclipsing system HW Vir in nearly 5:3 or 2:1 resonant captures and 
Beuermann et al. (2010) announced the existence of two planets in a 2:1 (or possibly 5:2) mean motion orbiting 
the post-common envelope binary NN Ser. Another interesting case is the W UMa-type binary star WZ Cep: Jeong \& Kim (2011) suggested 
that two periodicities of 41.3 yr and 11.8 yr exist in the $O$--$C$ residuals and indicate LITEs due to two circumbinary companions. 
The periods are exactly in a commensurable 7:2 relation between their mean motions. 

If the two circumbinary objects are on the main sequence and in the orbital plane ($i$=87$^\circ$.57) of the eclipsing pair SZ Her, 
the masses of the third and fourth bodies become $M_3$=0.22 M$_\odot$ and $M_4$=0.19 M$_\odot$, respectively. Following 
the empirical relations presented by Southworth (2009), the radii and temperatures are calculated to be $R_3$=0.23 R$_\odot$ and 
$T_3$=3018 K, and $R_4$=0.20 R$_\odot$ and $T_4$=3008 K for the third and fourth bodies, respectively. These values correspond 
to a spectral type of about M6--7 for both bodies and contribute only 0.1\% to the total bolometric luminosity of 
the supposed quadruple system. The semi-amplitudes of the systemic radial velocity variation of the eclipsing pair due to 
the additional objects are approximately 1 km s$^{-1}$. The two limits indicate that it will be difficult to detect these companions 
orbiting the eclipsing binary independently from spectroscopic data. This difficulty is further substantiated due to 
the large orbital periods suggested by the derived LITE models. However, the semi-major axes of the third and fourth companions 
relative to the binary center of mass are about 26.6 AU and 16.5 AU, respectively, corresponding to the angular sizes of 0.09 arcsec 
and 0.05 arcsec. The ($V-K$) color index for such M-type stars is about $+$7.3 mag, so the objects can be as bright as $K \sim$ 13 mag. 
Hence, they might be detected by careful observations with infrared photometry and direct speckle imaging interferometry.

In classical Algols, another possible mechanism for the period modulations is a magnetic activity cycle for systems with a secondary 
spectral type later than F5 (Hall 1989; Applegate 1992).  According to this mechanism, the variable rotational oblateness of a 
magnetically active star produces a change in its gravitational quadratic moment, hence forcing a change in the orbital period. 
With the periods and amplitudes for the two-LITE listed in Table 6, the model parameters were calculated for the secondary components 
from the Applegate formulae. The parameters are listed in Table 7, where the rms luminosity changes ($\Delta m_{\rm rms}$) converted 
to magnitude scale were obtained using equation (4) in the paper of Kim et al. (1997). The variations of 
the gravitational quadrupole moment ($\Delta Q$) are two orders of magnitude smaller than the typical values of $10^{51}-10^{52}$ 
for close binaries (Lanza \& Rodono 1999). Recently, Lanza (2006) noted that the Applegate mechanism is not sufficiently adequate 
to explain the period modulation of close binaries with a late-type secondary. These suggest that this kind of mechanism cannot 
explain the observed period variations of SZ Her.

Because SZ Her is in a semi-detached configuration with the less massive secondary component filling its inner Roche lobe, 
from both theoretical and intuitive viewpoints, a period increase could be produced through mass transfer from the secondary to 
the primary star. This implies that a long-term secular variation may be hidden in the $O$--$C$ data set and the Algol system 
may be in a weak phase of mass transfer. As listed in columns (5)--(6) of Table 6, fitting the eclipse timings to 
a quadratic {\it plus} two-LITE ephemeris indicates that the quadratic term ($Q$) represents a continuous period increase with 
a rate of d$P$/d$t$ = $+$2.5$\times$10$^{-10}$ d yr$^{-1}$. From these fits, it was found that this contribution is not significant 
and a secular term does not adequately describe the timing data, showing a larger $\chi^2$ value. Furthermore, this value 
corresponds to a mass transfer rate of 1.4$\times$10$^{-10}$ M$_\odot$ yr$^{-1}$, which is found to be the smallest rate amongst 
the classical semi-detached Algol-type systems. No difference in the final fitted parameters is observed when comparing the two-LITE 
($\chi^2$ = 1.013) and quadratic {\it plus} two-LITE ($\chi^2$ = 1.014) timing ephemeris. We are therefore left with 
the two-LITE ephemeris as a candidate that best explains the compiled timing data set of SZ Her with a possible mass transfer 
being negligible in this description. If the existence of the third and fourth components in SZ Her is true, they may have played 
an important role in the formation and evolution of the semi-detached eclipsing system, which may ultimately evolve into 
a contact configuration by re-distributing most of its angular momentum to the outer circumbinary companions. When more systematic 
and continuous observations (e.g., eclipse timings and spectroscopy) are undertaken, all of this is understood better than now and 
the absolute dimensions and evolutionary status of this system will be advanced greatly.

\acknowledgments{ }
The authors thank Professor Chun-Hwey Kim for his help using the $O$--$C$ database of eclipsing binaries and the staff of 
the Sobaeksan Optical Astronomy Observatory for assistance with our observations. We appreciate the careful reading and valuable comments 
of the anonymous referee and Dr. Tobias C. Hinse. This research has made use of the Simbad database maintained at CDS, Strasbourg, France. 
We have used data from the WASP public archive in this research. The WASP consortium comprises of the University of Cambridge, Keele University, 
University of Leicester, The Open University, The Queen's University Belfast, St. Andrews University and the Isaac Newton Group. Funding 
for WASP comes from the consortium universities and from the UK's Science and Technology Facilities Council.

\clearpage
\begin{figure}
 \includegraphics[]{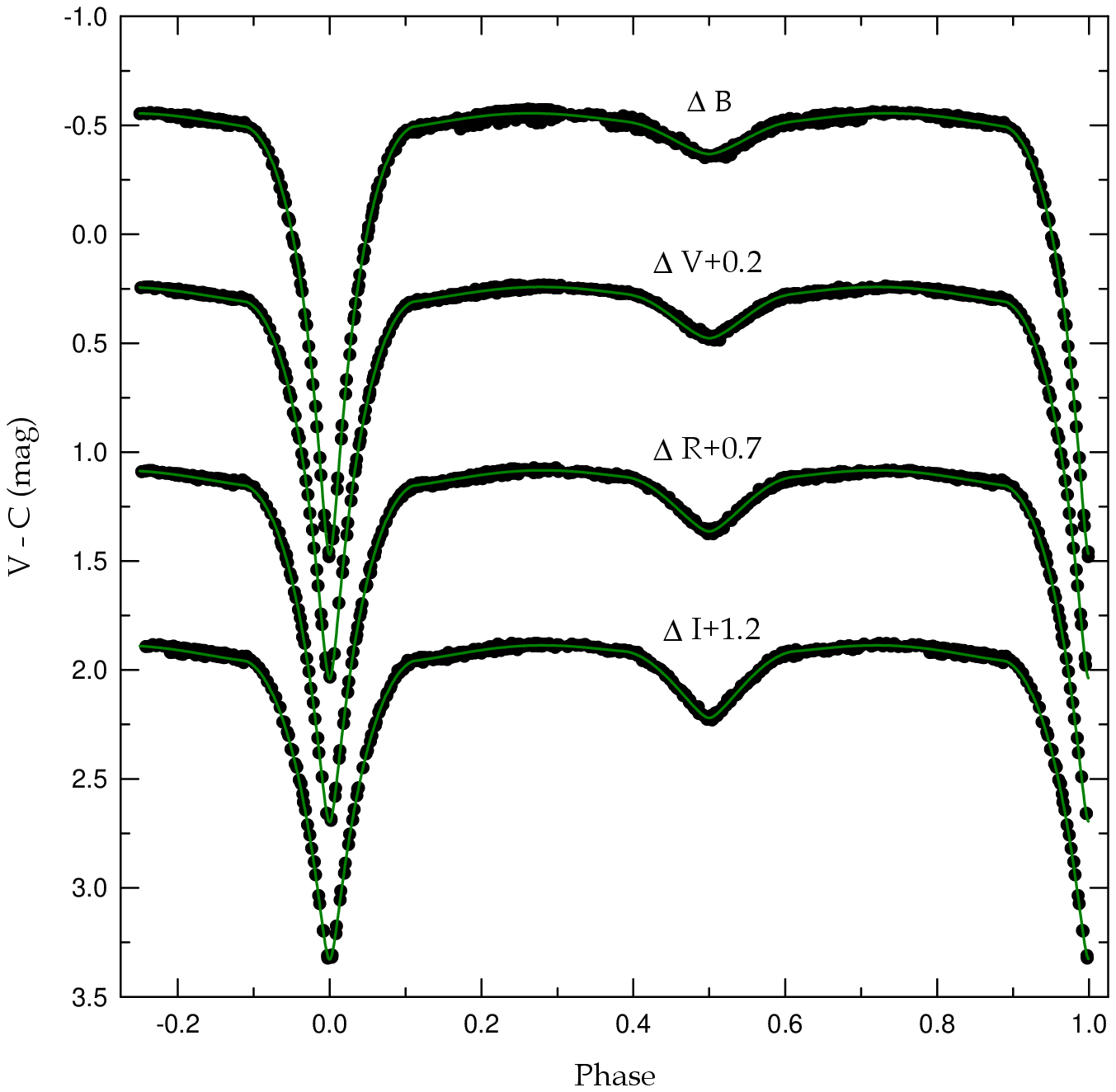}
 \caption{Light curves of SZ Her in the $B$, $V$, $R$, and $I$ bandpasses. Dots are individual measures and 
 solid lines represent the synthetic curves obtained from simultaneously fitting all SOAO data.}
 \label{Fig1}
\end{figure}

\begin{figure}
 \includegraphics[]{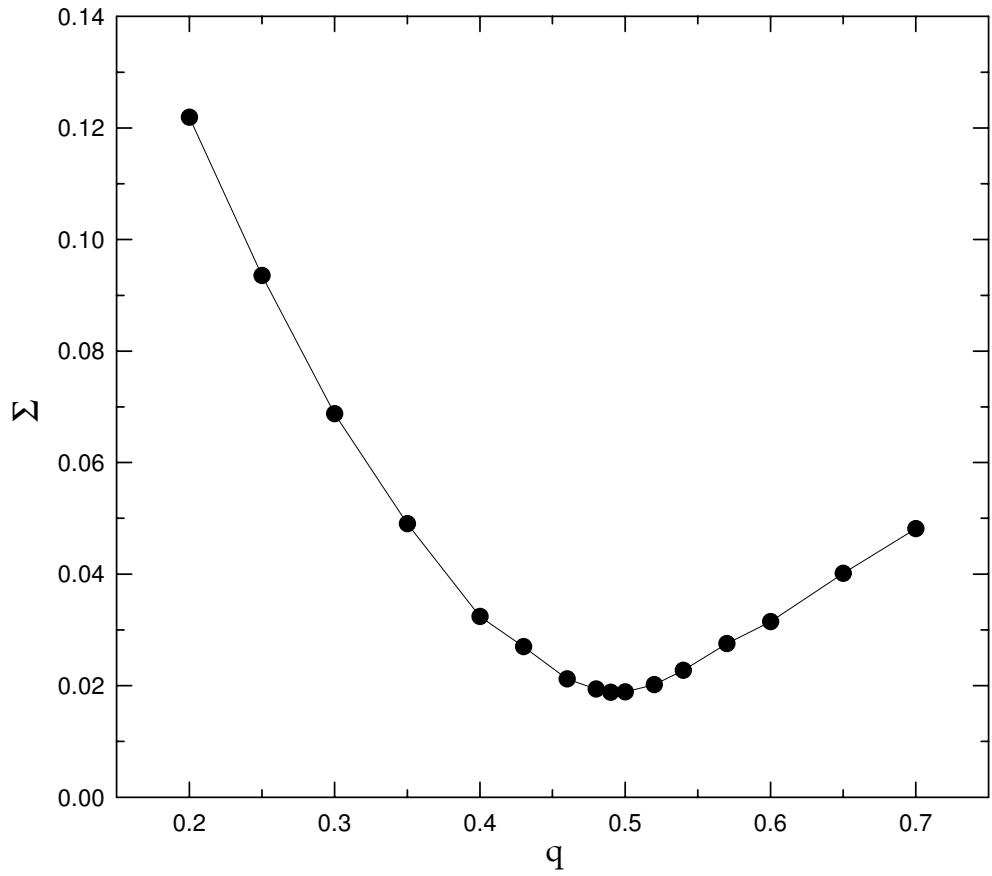}
 \caption{Behavior of $\Sigma$ as a function of mass ratio $q$, indicating a minimum value near $q$=0.49 for SZ Her.}
\label{Fig2}
\end{figure}

\begin{figure}
 \includegraphics[]{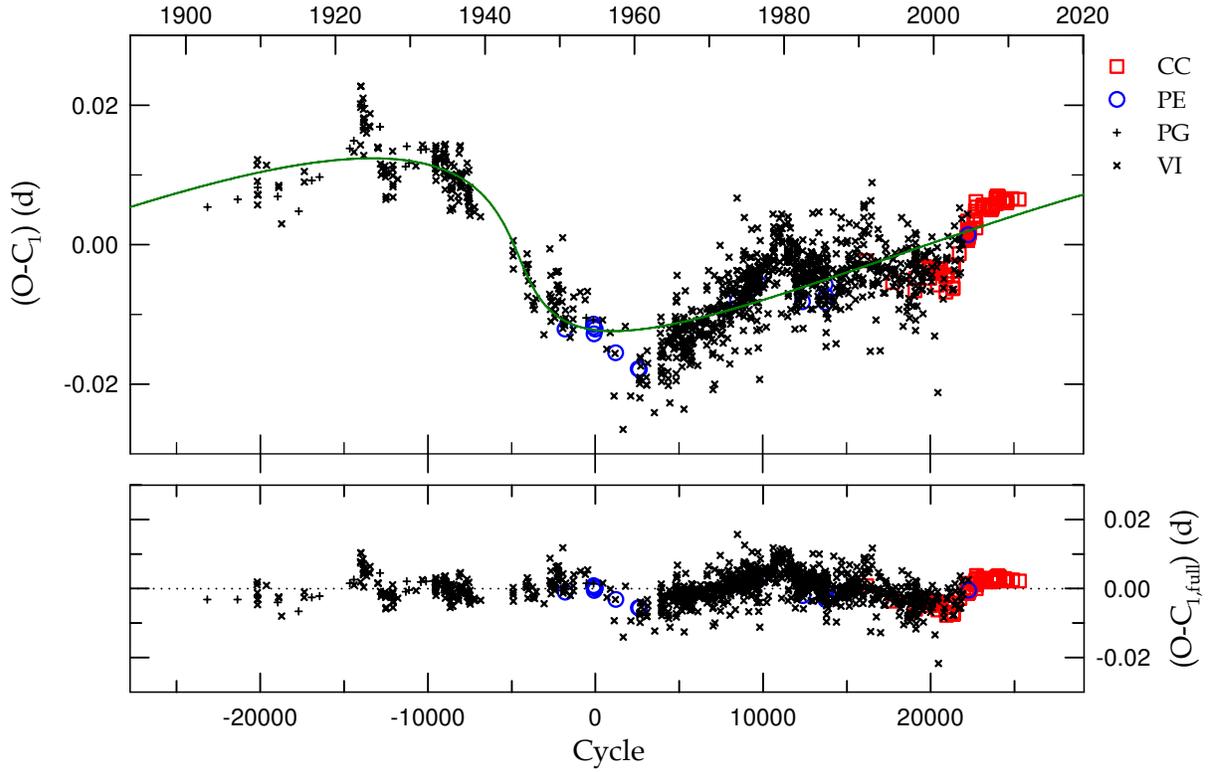}
 \caption{$O$--$C$ diagram of SZ Her. In the upper panel, constructed with the linear terms of equation (1), the continuous curve 
 represents the LITE orbit. The residuals from this LITE ephemeris are plotted in the lower panel where an additional short-term 
 oscillation appears to exist. CC, PE, PG, and VI denote CCD, photoelectric, photographic, and visual minima, respectively.}
 \label{Fig3}
\end{figure}

\begin{figure}
 \includegraphics[]{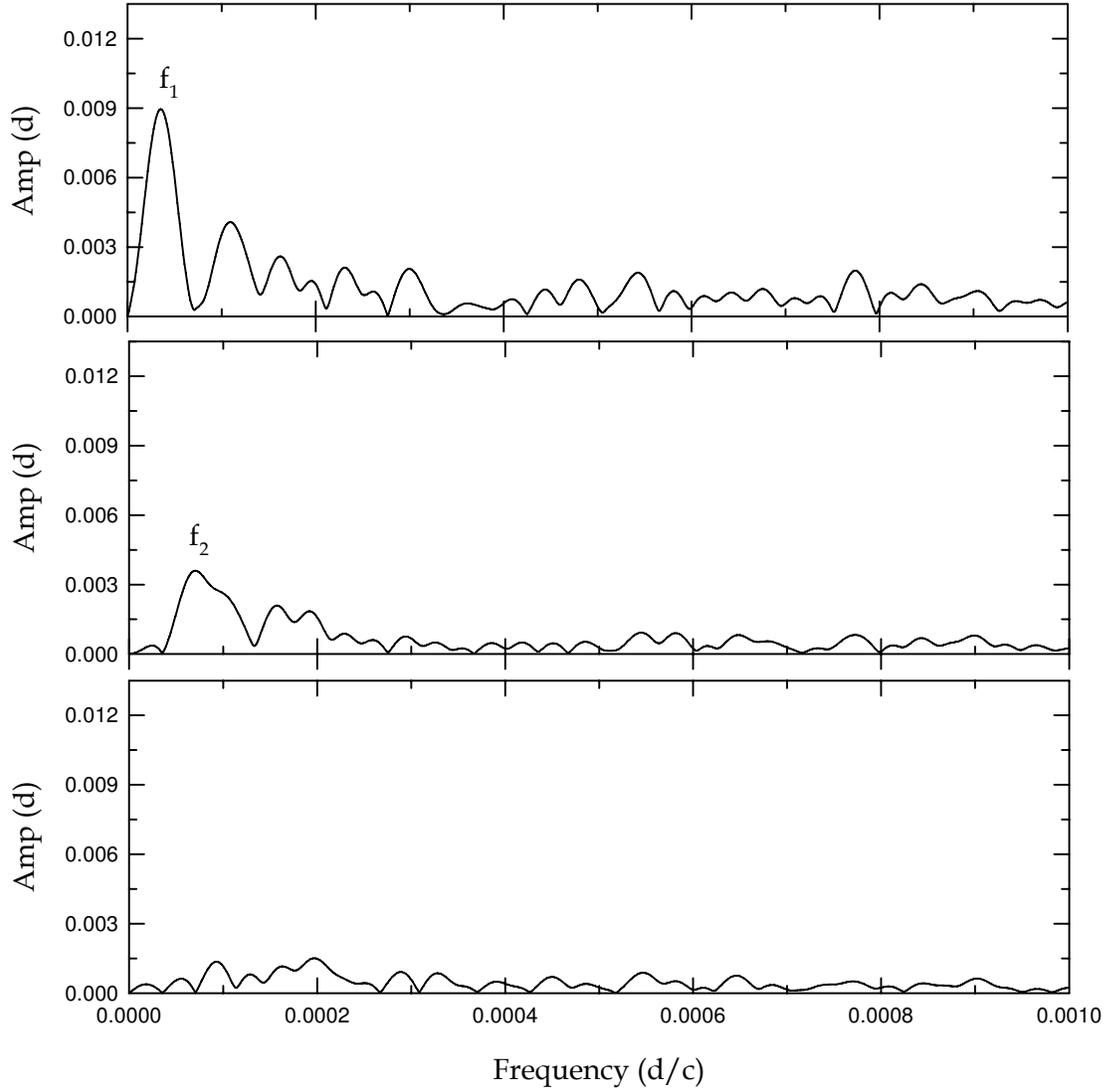}
 \caption{Periodogram from the PERIOD04 formalism for the $O$--$C_{1}$ residuals. As a result of the successive pre-whitening 
 procedures, two frequencies of $f_1$=0.0000353 cycle d$^{-1}$ and $f_2$=0.0000706 cycle d$^{-1}$ are detected and these 
 become periods of 77.6 and 38.8 yr.}
 \label{Fig4}
\end{figure}

\begin{figure}
 \includegraphics[]{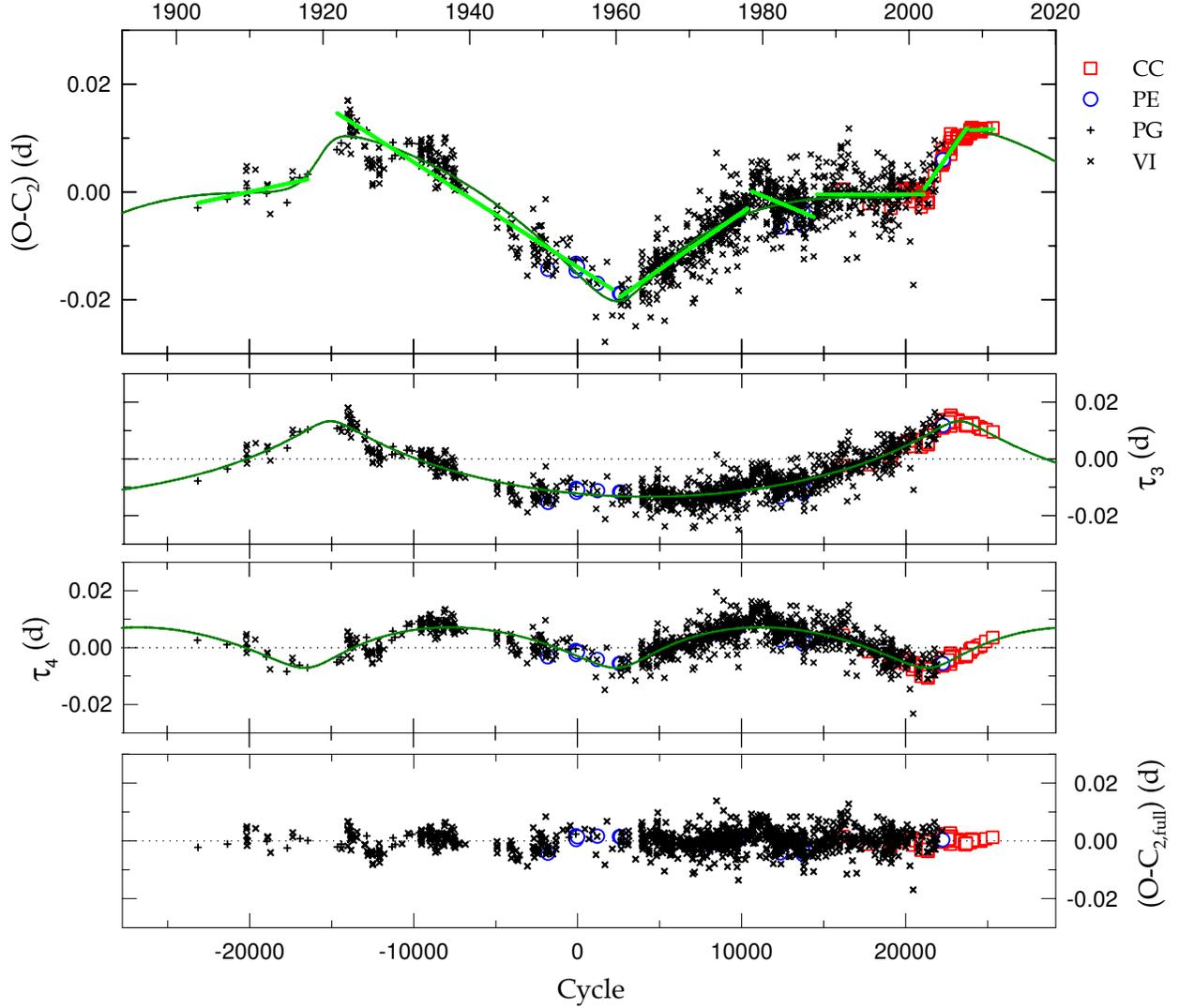}
 \caption{$O$--$C$ diagram of SZ Her with respect to equation (2). In the top panel, the two-LITE ephemeris is drawn as 
 the continuous curve and the straight lines represent the linear fits to the seven sections separated by period jumps. 
 The second and third panels display  the long- and short-term LITE orbits, respectively. The bottom panel shows the residuals 
 from the complete ephemeris, which demonstrates much better performance than that in Figure 3. } 
\label{Fig5}
\end{figure}

\clearpage
\begin{deluxetable}{lccccc}
\tablewidth{0pt} 
\tablecaption{Coordinates and Photometric Data for the Program Stars.}
\tablehead{
\colhead{Star} & \colhead{GSC}  & \colhead{RA (J2000)}                & \colhead{DEC (J2000)}                 & $V\rm_T\rm ^a$   &  $(B-V)\rm_T\rm ^a$}             
\startdata                                                                                                                             
SZ Herculis    &  2610-0129     &  17$^{\rm h}$39$^{\rm m}$36$\fs81$  &  +32$^{\circ}$56${\rm '}$46$\farcs$7  & $+$10.47         &  $+$0.38                 \\
Comparison     &  2610-1116     &  17$^{\rm h}$39$^{\rm m}$00$\fs20$  &  +32$^{\circ}$54${\rm '}$36$\farcs$0  & $+$11.19         &  $+$1.28                 \\
Check          &  2610-0821     &  17$^{\rm h}$39$^{\rm m}$03$\fs40$  &  +32$^{\circ}$58${\rm '}$56$\farcs$7  &  --              &  --                      \\
\enddata
\tablenotetext{a}{From the Tycho-2 Catalogue (H\o g et al. 2000).}
\end{deluxetable}

\begin{deluxetable}{crcrcrcr}
\tabletypesize{\small}
\tablewidth{0pt} 
\tablecaption{CCD Photometric Observations of SZ Her}
\tablehead{
\colhead{HJD} & \colhead{$\Delta B$} & \colhead{HJD} & \colhead{$\Delta V$} & \colhead{HJD} & \colhead{$\Delta R$} & \colhead{HJD} & \colhead{$\Delta I$}
}
\startdata
2,454,525.24638 & -0.389  &  2,454,525.24769 & 0.183  &  2,454,525.24881 & 0.519  &  2,454,525.25950 & 0.756   \\
2,454,525.25125 & -0.428  &  2,454,525.25268 & 0.152  &  2,454,525.25380 & 0.504  &  2,454,525.26414 & 0.759   \\
2,454,525.25615 & -0.456  &  2,454,525.25753 & 0.128  &  2,454,525.25859 & 0.478  &  2,454,525.26878 & 0.754   \\
2,454,525.26079 & -0.469  &  2,454,525.26217 & 0.117  &  2,454,525.26323 & 0.473  &  2,454,525.27343 & 0.742   \\
2,454,525.26543 & -0.469  &  2,454,525.26681 & 0.120  &  2,454,525.26787 & 0.457  &  2,454,525.27795 & 0.743   \\
2,454,525.27007 & -0.468  &  2,454,525.27145 & 0.117  &  2,454,525.27251 & 0.457  &  2,454,525.28251 & 0.748   \\
2,454,525.27472 & -0.479  &  2,454,525.27606 & 0.101  &  2,454,525.27705 & 0.445  &  2,454,525.28704 & 0.753   \\
2,454,525.27926 & -0.488  &  2,454,525.28061 & 0.111  &  2,454,525.28160 & 0.449  &  2,454,525.29158 & 0.741   \\
2,454,525.28381 & -0.486  &  2,454,525.28516 & 0.103  &  2,454,525.28615 & 0.456  &  2,454,525.29601 & 0.751   \\
2,454,525.28835 & -0.498  &  2,454,525.28970 & 0.100  &  2,454,525.29069 & 0.448  &  2,454,525.30726 & 0.741   \\
\enddata
\tablecomments{This table is available in its entirety in machine-readable and Virtual Observatory (VO) forms in the online journal. 
A portion is shown here for guidance regarding its form and content.}
\end{deluxetable}

\begin{deluxetable}{lcc}
\tablewidth{0pt} 
\tablecaption{Photometric Solutions of SZ Her.}
\tablehead{
\colhead{Parameter}          & \colhead{Primary} & \colhead{Secondary}                                                       
}
\startdata                                                                                                                   
$T_0$ (HJD)                  & \multicolumn{2}{c}{2,454,588.167997$\pm$0.000026}     \\
$P$ (d)                      & \multicolumn{2}{c}{0.8180837$\pm$0.0000034}           \\
$q$                          & \multicolumn{2}{c}{0.4715$\pm$0.0032}                 \\
$i$ ($^\circ$)               & \multicolumn{2}{c}{87.566$\pm$0.078}                  \\
$T$ (K)                      & 7,262$\pm$78             & 4,881$\pm$33               \\
$\Omega$                     & 3.668$\pm$0.011          & 2.8207                     \\
$A$                          & 1.0                      & 0.928$\pm$0.031            \\
$g$                          & 1.0                      & 0.253$\pm$0.053            \\
$X$                          & 0.468                    & 0.534                      \\
$x_{B}$                      & 0.661$\pm$0.026          & 0.738$\pm$0.112            \\
$x_{V}$                      & 0.549$\pm$0.029          & 0.865$\pm$0.079            \\
$x_{R}$                      & 0.464$\pm$0.031          & 0.706$\pm$0.067            \\
$x_{I}$                      & 0.361$\pm$0.033          & 0.466$\pm$0.061            \\
$L/(L_1+L_2)_{B}$            & 0.9168$\pm$0.0034        & 0.0832                     \\
$L/(L_1+L_2)_{V}$            & 0.8738$\pm$0.0028        & 0.1262                     \\
$L/(L_1+L_2)_{R}$            & 0.8316$\pm$0.0024        & 0.1684                     \\
$L/(L_1+L_2)_{I}$            & 0.7861$\pm$0.0023        & 0.2139                     \\
$r$ (pole)                   & 0.3107$\pm$0.0011        & 0.2952$\pm$0.0005          \\
$r$ (point)                  & 0.3288$\pm$0.0015        & 0.4234$\pm$0.0022          \\
$r$ (side)                   & 0.3180$\pm$0.0012        & 0.3080$\pm$0.0006          \\
$r$ (back)                   & 0.3245$\pm$0.0013        & 0.3405$\pm$0.0006          \\
$r$ (volume)$\rm ^a$         & 0.3180                   & 0.3159                     
\tablenotetext{a}{Mean volume radius.}
\enddata
\end{deluxetable}

\begin{deluxetable}{lcc}
\tablewidth{0pt} 
\tablecaption{Absolute Parameters for SZ Her.}
\tablehead{
\colhead{Parameter}              & \colhead{Primary} & \colhead{Secondary}
}
\startdata
$M$/M$_\odot$                    &  1.58             &  0.75            \\
$R$/R$_\odot$                    &  1.55             &  1.54            \\
$\log$ $g$ (cgs)                 &  4.26             &  3.94            \\
$\log$ $\rho$/$\rho_\odot$       &  $-$0.37          &  $-$0.69         \\
$T$ (K)                          &  7,262            &  4,881           \\
$L$/L$_\odot$                    &  5.98             &  1.20            \\
$M_{\rm bol}$ (mag)              &  $+$2.79          &  $+$4.53         \\
BC (mag)                         &  $+$0.03          &  $-$0.36         \\
$M_{V}$ (mag)                    &  $+$2.76          &  $+$4.89         \\
$M_{\rm V,total}\rm ^a$ (mag)    &  \multicolumn{2}{c}{$+$2.61}         \\
Distance (pc)                    &  \multicolumn{2}{c}{294}             \\
\enddata
\tablenotetext{a}{Absolute visual magnitude from both components.}
\end{deluxetable}

\begin{deluxetable}{llccllc}
\tablewidth{0pt}
\tablecaption{New Times of Minimum Light for SZ Her.}
\tablehead{
\colhead{HJD} & \colhead{Error} & \colhead{Min} && \colhead{HJD} & \colhead{Error} & \colhead{Min}  \\
\colhead{(2,450,000+)} & & && \colhead{(2,450,000+)} & & }
\startdata
3,128.67869            &  $\pm$0.00006  & I   &&   3,237.48623            &  $\pm$0.00014  & I      \\
3,137.67757            &  $\pm$0.00007  & I   &&   3,242.39511            &  $\pm$0.00007  & I      \\
3,142.58640            &  $\pm$0.00004  & I   &&   3,246.48534            &  $\pm$0.00015  & I      \\
3,144.63307            &  $\pm$0.00040  & II  &&   3,253.43933            &  $\pm$0.00048  & II     \\
3,151.58572            &  $\pm$0.00006  & I   &&   3,262.43879            &  $\pm$0.00066  & II     \\
3,153.63084            &  $\pm$0.00022  & II  &&   3,278.39164            &  $\pm$0.00046  & I      \\
3,155.67585            &  $\pm$0.00011  & I   &&   3,855.55946            &  $\pm$0.00061  & II     \\
3,156.49399            &  $\pm$0.00006  & I   &&   3,882.55882            &  $\pm$0.00027  & II     \\
3,158.54003            &  $\pm$0.00035  & II  &&   3,907.51075            &  $\pm$0.00005  & I      \\
3,160.58427            &  $\pm$0.00015  & I   &&   3,921.41866            &  $\pm$0.00008  & I      \\
3,162.62941            &  $\pm$0.00036  & II  &&   3,923.46831            &  $\pm$0.00040  & II     \\
3,164.67494            &  $\pm$0.00003  & I   &&   3,943.50694            &  $\pm$0.00004  & I      \\
3,165.49276            &  $\pm$0.00006  & I   &&   3,948.41539            &  $\pm$0.00016  & I      \\
3,167.53872            &  $\pm$0.00022  & II  &&   4,298.56066            &  $\pm$0.00014  & I      \\
3,169.58367            &  $\pm$0.00005  & I   &&   4,303.46897            &  $\pm$0.00007  & I      \\
3,171.63073            &  $\pm$0.00033  & II  &&   4,307.55961            &  $\pm$0.00022  & I      \\
3,173.67431            &  $\pm$0.00002  & I   &&   4,312.46786            &  $\pm$0.00018  & I      \\
3,174.49230            &  $\pm$0.00005  & I   &&   4,328.42428            &  $\pm$0.00054  & I      \\
3,176.53844            &  $\pm$0.00051  & II  &&   4,330.46611            &  $\pm$0.00014  & I      \\
3,178.58273            &  $\pm$0.00004  & I   &&   4,581.21434$\rm ^a$    &  $\pm$0.00037  & II     \\
3,180.62827            &  $\pm$0.00050  & II  &&   4,581.62260            &  $\pm$0.00012  & I      \\
3,183.08448$\rm ^a$    &  $\pm$0.00063  & II  &&   4,588.16795$\rm ^a$    &  $\pm$0.00006  & I      \\
3,183.49127            &  $\pm$0.00003  & I   &&   4,592.25831$\rm ^a$    &  $\pm$0.00006  & I      \\
3,185.53634            &  $\pm$0.00043  & II  &&   4,604.12114$\rm ^a$    &  $\pm$0.00042  & II     \\
3,190.44593            &  $\pm$0.00043  & II  &&   4,631.52689            &  $\pm$0.00004  & I      \\
3,192.49023            &  $\pm$0.00007  & I   &&   4,638.48360            &  $\pm$0.00050  & II     \\
3,194.53696            &  $\pm$0.00043  & II  &&   4,640.52602            &  $\pm$0.00009  & I      \\
3,196.58119            &  $\pm$0.00005  & I   &&   4,644.61676            &  $\pm$0.00007  & I      \\
3,199.44569            &  $\pm$0.00031  & II  &&   4,645.43473            &  $\pm$0.00003  & I      \\
3,201.48976            &  $\pm$0.00004  & I   &&   4,647.48195            &  $\pm$0.00048  & II     \\
3,203.53628            &  $\pm$0.00047  & II  &&   4,651.57417            &  $\pm$0.00057  & II     \\
3,205.58026            &  $\pm$0.00005  & I   &&   4,663.43290            &  $\pm$0.00010  & I      \\
3,208.44389            &  $\pm$0.00057  & II  &&   4,669.57011            &  $\pm$0.00097  & II     \\
3,219.48818            &  $\pm$0.00008  & I   &&   4,672.43205            &  $\pm$0.00006  & I      \\
3,224.39681            &  $\pm$0.00011  & I   &&   4,674.48004            &  $\pm$0.00052  & II     \\
3,226.44262            &  $\pm$0.00034  & II  &&   4,681.43086            &  $\pm$0.00005  & I      \\
3,228.48723            &  $\pm$0.00007  & I   &&   4,683.47803            &  $\pm$0.00027  & II     \\
3,230.53328            &  $\pm$0.00062  & II  &&   4,685.52134            &  $\pm$0.00008  & I      \\
3,233.39568            &  $\pm$0.00031  & I   &&   5,685.23465$\rm ^a$    &  $\pm$0.00006  & I      \\
3,235.44175            &  $\pm$0.00042  & II  &&                          &                &        \\
\enddata
\tablenotetext{a}{SOAO minimum epochs. The others from the WASP public data.}
\end{deluxetable}

\begin{deluxetable}{lcccccccc}
\tabletypesize{\scriptsize}
\tablewidth{0pt}
\tablecaption{Parameters for the LITE Orbits of SZ Her.}
\tablehead{
\colhead{Parameter}      & \colhead{Single-LITE}  && \multicolumn{2}{c}{Two-LITE}                       && \multicolumn{2}{c}{Quadratic {\it plus} Two-LITE}     & \colhead{Unit}         \\ [1.5mm] \cline{4-5} \cline{7-8}\\ [-2.0ex]
\colhead{}               & \colhead{$\tau_{3}$}   && \colhead{$\tau_{3}$}       & \colhead{$\tau_{4}$}  && \colhead{$\tau_{3}$}       & \colhead{$\tau_{4}$}     &                         
}                                                                                                                                                                                         
\startdata                                                                                                                                                                                
$T_0$                    &  2,434,987.3975(15)    &&  \multicolumn{2}{c}{2,434,987.39930(78)}           &&  \multicolumn{2}{c}{2,434,987.39925(75)}              &   HJD                  \\
$P$                      &  0.818096071(92)       &&  \multicolumn{2}{c}{0.818095789(46)}               &&  \multicolumn{2}{c}{0.818095788(44)}                  &   d                    \\
$a_{12}\sin i_{3,4}$     &  3.23(66)              &&  2.31(18)                  &  1.24(20)             &&  2.30(17)                  &  1.24(19)                &   au                   \\
$\omega$                 &  187.3(5.7)            &&  88.6(7.5)                 &  285(10)              &&  88.4(7.2)                 &  285(9)                  &   deg                  \\
$e$                      &  0.75(10)              &&  0.718(90)                 &  0.48(17)             &&  0.720(87)                 &  0.48(16)                &                        \\
$n$                      &  0.00809(78)           &&  0.01148(13)               &  0.02320(17)          &&  0.01148(12)               &  0.02319(17)             &   deg d$^{-1}$         \\
$T$                      &  2,431,348(1271)       &&  2,422,631(312)            &  2,406,158(320)       &&  2,422,629(299)            &  2,406,141(320)          &   HJD                  \\
$P_{3,4}$                &  122(12)               &&  85.8(1.0)                 &  42.5(1.1)            &&  85.8(0.9)                 &  42.5(0.3)               &   yr                   \\
$K$                      &  0.0124(26)            &&  0.0133(10)                &  0.0071(11)           &&  0.0133(10)                &  0.0071(11)              &   d                    \\
$f(M_{3,4})$             &  0.00226(51)           &&  0.00167(13)               &  0.00106(17)          &&  0.00166(12)               &  0.00106(16)             &   M$_\odot$            \\
$M_{3,4} \sin i_{3,4}$   &  0.247(29)             &&  0.222(9)                  &  0.189(15)            &&  0.221(8)                  &  0.189(14)               &   M$_\odot$            \\[0.5mm]
$Q$                      &                        &&                            &                       && \multicolumn{2}{c}{+(0.28$\pm$2.10)$\times 10^{-12}$} &  d                     \\
$dP$/$dt$                &                        &&                            &                       && \multicolumn{2}{c}{+(0.25$\pm$1.87)$\times 10^{-9}$}  &  d yr$^{-1}$           \\[0.5mm]
Reduced $\chi^2$         &  1.663                 &&  \multicolumn{2}{c}{1.013}                         &&  \multicolumn{2}{c}{1.014}                            &                        \\
\enddata
\end{deluxetable}

\begin{deluxetable}{lccc}
\tablewidth{0pt}
\tablecaption{Applegate Parameters for the Cool Secondary of SZ Her.}
\tablehead{
\colhead{Parameter}       & \colhead{Long-term}    & \colhead{Short-term}    & \colhead{Unit}
}
\startdata
$\Delta P$                & 0.1885                 &  0.2031                 &  s                     \\
$\Delta P/P$              & $2.67\times10^{-6}$    &  $2.87\times10^{-6}$    &                        \\
$\Delta Q$                & ${5.08\times10^{49}}$  &  ${5.48\times10^{49}}$  &  g cm$^2$              \\
$\Delta J$                & ${1.39\times10^{47}}$  &  ${1.49\times10^{47}}$  &  g cm$^{2}$ s$^{-1}$   \\
$I_{\rm s}$               & ${1.14\times10^{54}}$  &  ${1.14\times10^{54}}$  &  g cm$^{2}$            \\
$\Delta \Omega$           & ${1.21\times10^{-7}}$  &  ${1.31\times10^{-7}}$  &  s$^{-1}$              \\
$\Delta \Omega / \Omega$  & ${1.36\times10^{-3}}$  &  ${1.47\times10^{-3}}$  &                        \\
$\Delta E$                & ${3.36\times10^{40}}$  &  ${3.90\times10^{40}}$  &  erg                   \\
$\Delta L_{\rm rms}$      & ${3.90\times10^{31}}$  &  ${9.15\times10^{31}}$  &  erg s$^{-1}$          \\
                          & 0.0100                 &  0.0235                 &  L$_\odot$             \\
                          & 0.0083                 &  0.0195                 &  $L_{\rm p,s}$         \\
$\Delta m_{\rm rms}$      & $\pm$0.0015            &  $\pm$0.0035            &  mag                   \\
$B$                       & 2.9                    &  4.2                    &  kG                    \\
\enddata
\end{deluxetable}

\end{document}